\documentclass[draftclsnofoot, 12pt,onecolumn]{IEEEtran}
\usepackage{amsmath}
\usepackage{subfigure}
\usepackage{graphicx}
\usepackage{latexsym}
\usepackage{amssymb,amsbsy,verbatim,array}
\usepackage{epsfig}
\usepackage{cite}
\usepackage{color}
\usepackage{soul}

\usepackage{calc}

\newtheorem{theorem}{Theorem}

\newcommand{\bit}{\begin{itemize}}
\newcommand{\eit}{\end{itemize}}

\newcommand{\bc}{\begin{center}}
\newcommand{\ec}{\end{center}}

\newcommand{\ba}{\begin{array}}
\newcommand{\ea}{\end{array}}

\newcommand{\beq}{\begin{equation}}
\newcommand{\eeq}{\end{equation}}

\newcommand{\beqn}{\begin{equation*}}
\newcommand{\eeqn}{\end{equation*}}

\newcommand{\bean}{\begin{eqnarray*}}
\newcommand{\eean}{\end{eqnarray*}}
\newcommand{\bea}{\begin{eqnarray}}
\newcommand{\eea}{\end{eqnarray}}

\newcommand{\Cc}{{\mathcal C}}

\newcommand{\Qc}{{\mathcal Q}}

\newcommand{\Vc}{{\mathcal V}}

\def\pl{{p_{s,out}}}
\def\po{{p_{c,out}}}
\def\poAF{{p_{c,out}^{AF}}}
\def\polb{{p_{c,out}^{lb}}}

\begin{document}
\title{Near-Optimal Modulo-and-Forward Scheme for the Untrusted Relay Channel
\thanks{
S. Zhang is with the School of Information Engineering,
Shenzhen University, Shenzhen, China (e-mail: {\tt zsl@szu.edu.cn}),
and also with the Electrical Engineering Department, Stanford
CA, USA. L. Fan is with the Department of Electronic
Engineering, Shantou University, Shantou, China (e-mail: {\tt
lsfan@stu.edu.cn}). M. Peng is with the Beijing University of Posts
\text{\&} Telecommunications, Beijing, China(e-mail: {\tt
pmg@bupt.edu.cn}), and also with the School of Engineering and
Applied Science, Princeton University, Princeton, NJ, USA. H. V.
Poor is with the School of Engineering and Applied Science,
Princeton University, Princeton, NJ, USA (e-mail: {\tt
poor@princeton.edu}).
This research was partially supported in part by the U. S. National Science Foundation
under Grant CMMI-1435778.
}}
\author{\IEEEauthorblockN{Shengli Zhang, Lisheng Fan, Mugen Peng, and H. Vincent Poor}}

%
%\author{\authorblockN{Shengli Zhang$^{1}$,
%Lisheng Fan$^2$, Mugen Peng$^3$, and H. Vincent Poor $^4$}\\
%%\vspace*{-0.4cm}
%
%\authorblockA{$^1$ School of Information Engineering, Shenzhen University, Shenzhen, China, Email: zsl@szu.edu.cn} \\
%
%\authorblockA{$^2$ Department of Electronic
%Engineering, Shantou University, Shantou, China. Email:
%lsfan@stu.edu.cn} \\
%
%\authorblockA{$^3$ Key Laboratory of Universal Wireless Communication, Ministry
%of Education, Beijing University of Posts and Telecommunications, Beijing, China. Email:
%pmg@bupt.edu.cn} \\
%
%\authorblockA{$^4$ Department of Electrical Engineering, Princeton
%University, Princeton, NJ, USA. Email: poor@princeton.edu}
%}

\pagestyle{headings} \maketitle \thispagestyle{empty}

\begin{abstract}
%\boldmath
This paper studies an untrusted relay channel, in which the
destination sends artificial noise simultaneously with the source
sending a message to the relay, in order to protect the source's
confidential message. The traditional amplify-and-forward (AF)
scheme shows poor performance in this situation because of the
interference power dilemma: providing better security by using
stronger artificial noise will  decrease the confidential
message power from the relay to the destination. To solve this
problem, a modulo-and-forward (MF) operation at the relay with
nested lattice encoding at the source is proposed. For this system
with full channel state information at the transmitter (CSIT),
theoretical analysis shows that the proposed MF scheme approaches
the secrecy capacity within $1/2$ bit for any channel realization,
and hence achieves full generalized security degrees of freedom
(G-SDoF). In contrast,  the AF scheme can only
achieve a small fraction of the G-SDoF. For this system without any
CSIT, the total outage event, defined as either connection outage or
secrecy outage, is introduced. Based on this total outage
definition, analysis shows that the proposed MF scheme  achieves
the full generalized secure diversity gain (G-SDG) of order one. On
the other hand, the AF
 scheme can only achieve a G-SDG of $1/2$ at most.
\end{abstract}

\clearpage

\section{Introduction}

The broadcast nature of wireless transmission creates significant
security concerns, and physical layer techniques can be used in part
to address these concerns. The theoretical basis for physical-layer
security can be traced back to Shannon's work on perfect secrecy
\cite{shannon_secrecy:48}, and to subsequent work by Wyner
\cite{wyner_secrecy:75}, Leung et al. \cite{leung_secrecy:78}, and
others on the wire-tap channel. The basic idea of physical-layer
security is to exploit the destination's advantages (e.g, better
channel quality) over the eavesdropper. More recent works have
investigated this problem in fading channels, including analyses of
the fading secrecy capacity \cite{parada_secure_capacity_fading:05,
gopala_secure_capacity_fading:08} and the secrecy outage probability
\cite{barros_secure_outage:06, tang_connection_outage_2009}. On the
other hand, new coding and modulation schemes have been proposed to
achieve physical-layer security, including Low Density Parity Check
(LDPC) codes in \cite{Thangaraj_secure_ldpc:07} and
\cite{klinc_secure_ldpc:11}, nested lattice codes in
\cite{choo_nested_lattice:11} and \cite{ he_nested_lattice:08}, and
nested polar codes in \cite{Mahdavifar_secure_polar:11} and
\cite{andersson_secure_polar:10}.

In recent years, several extended models of the wire-tap channel
have been studied and one of these is the untrusted relay channel.
In the untrusted relay channel, the source relies on a relay node to
forward information to the destination, while keeping the
transmitted information confidential from the relay. An example is
the two-way untrusted relay channel with two-phase physical-layer
network coding \cite{mobicom:06}, in which the superimposed signals
at the relay protect each other's information with minimal rate loss
compared to capacity \cite{lu_pnc_security:09,
Mukherjee_anc_security:10}. Similar ideas have been applied in the
one-way untrusted relay channel in which the destination artificially
transmits some interference to the relay while the source is
transmitting. Depending on the processing at the relay,  such schemes can be classified into two categories. The first category
is amplify-and-forward (AF), in which the relay simply amplifies the
received signal under its power constraint and then forwards it to
the destination \cite{huang_secure_fading:13, he_untrusted_relay:08,
song_secure_fading:12}. Although AF is simple to implement, its
performance is severely limited by the interference power dilemma:
more power of the relay is wasted on forwarding the interfering
signal (which is useless to the destination) when the destination
transmits with larger power to protect the confidential message;
alternatively the confidential message is less well protected when
the destination transmits with smaller power. The other category is
decode-and-forward (DF) \cite{he_lattice:13}, in which both source
signal and interference signal are encoded by lattice codes and
arrive at the relay in perfect synchrony\footnote{Perfect
synchronization here refers to the synchronization of signal amplitude,
carrier frequency and carrier phase. }, followed by channel decoding
to obtain the noiseless network code produced by this signal. The
DF scheme performs better than the AF scheme in the high SNR region
but performs worse in the low SNR region, despite the cost of
perfect synchronization. Schemes extending these ideas to multiple
channels and fading channels can be found in many works, such as
\cite{bloch_security:08} and \cite{ Jeong_mimo_security:12}.

To counter the shortcomings of these existing techniques and inspired by the modulo-lattice
additive noise (MLAN) channel \cite{EZ:04} in which modulo processing
at the receiver loses very little information, here we propose a
novel modulo-and-forward (MF) scheme at the relay. In this scheme,
the confidential message from the source node is encoded with a
nested lattice code \cite{choo_nested_lattice:11} while the
artificial interference message from the destination is Gaussian.
When the two messages arrive at the relay node simultaneously, the
relay maps the superimposed signal to a new signal with a modulo
operation according to the source lattice. As a result, the total
power of the new message is reduced to that of the source lattice,
with almost no loss of useful information. In this way, the proposed
MF scheme solves the interference power dilemma of the AF scheme by
relaying a signal with only the source power, without relation to
the interference power. Moreover, MF does not require perfect
synchronization as in the DF scheme.

We analyze the secure performance of this MF scheme for two
different cases. For Gaussian channels with full channel state
information at the transmitter (CSIT), our analysis shows that the MF
scheme  approaches the secrecy capacity of the untrusted relay
channel within 1/2 bit for all channel realizations; hence it achieves
 the full generalized secure degrees of freedom (G-SDoF),
while the AF scheme can achieve only a small fraction of the
G-SDoF. The achievable secrecy rate of MF is also
better than that of DF, which as noted above requires fine %/ {\color{green}requires a stricter}
synchronization.

For fading channels without any CSIT, we
characterize the total outage probability, which includes the connection
outage probability at the destination and the secrecy outage
probability at the relay node \cite{tang_connection_outage_2009}.
Beyond achieving a smaller outage probability as expected, the MF
scheme  achieves essential improvement over the AF scheme.
Defining the generalized secure diversity as the rate with which the
total outage probability goes to zero in the high SNR region, our
analysis shows that the MF scheme achieves full diversity gain of
1 as long as the ratio of the destination signal power to the
source signal power goes to infinity. The AF scheme, however, can
only achieve a diversity gain of 1/2 at most, due to the
interference power dilemma.

The contributions of this paper are as follows.
\begin{enumerate}
\item \textbf{The MF Scheme:} We propose the lattice code based modulo-and-forward scheme for the untrusted relay channel. This scheme is of practical interest since it only needs symbol level time synchronization and low complexity processing to obtain much better performance than other schemes.

\item \textbf{Analysis of the Secure Capacity and G-SDoF in Gaussian Channels:}
In Gaussian channels with full CSIT, we prove that the MF scheme
 approaches the secrecy capacity within one-half bit for any channel
realization. Hence, the MF scheme achieves the full
G-SDoF, while the AF scheme
only achieves a very small fraction of the G-SDoF.

\item \textbf{Analysis of the Secure Diversity in Fading Channels:} In fading channels without  CSIT, the MF scheme achieves the full generalized secure diversity gain (G-SDG) value of 1, while the AF scheme only achieves a G-SDG of 1/2 due to the interference power dilemma.
\end{enumerate}

The remainder of this paper is organized as follows. Section II
presents the detailed untrusted relay model and the notation used
therein. For the proposed modulo-and-forward scheme, Section III elaborates the encoding process at the source
node, the forwarding process at the relay node and the decoding
process at the destination node.
%Section IV provides the performance
%metric of secure communication under different conditions of CSI at
%the transmitter.
Under the assumption of full CSIT, Section IV analyzes the achievable rate and the G-SDoF
of the MF scheme, with a comparison to the AF and DF schemes.
Under the assumption of no CSIT, Section V
analyzes the connection outage and secrecy outage probabilities, as well as the system secure
diversity gain. Finally, Section VI concludes the paper.

\section{System Model}

\begin{figure}[tbp]
\begin{center}
\includegraphics[width=5.0in]{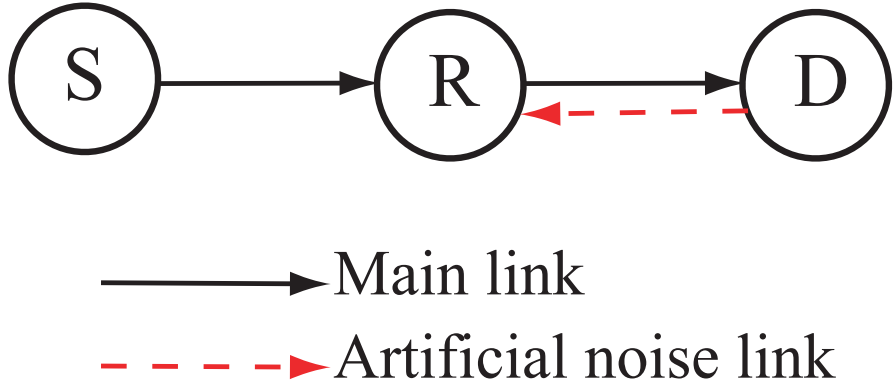}
\caption{System model for the untrusted relay channel based on lattice coding.}
\label{system_model}
\end{center}
\end{figure}

Fig. \ref{system_model} shows the system model for the considered
untrusted relay channel, where the source $S$ needs to transmit a
confidential message to the destination $D$ with $R$ working as a
relay. All nodes are equipped with single antennas, and operate
in a time-division half-duplex mode. In this setting, although the
source relies on the relay to forward the message, it does not trust
the relay and would like to keep the confidential information secret
from the relay. This model can find application in many practical
scenarios such as renting a satellite from a third party
to relay confidential messages.

To prevent eavesdropping at $R$, an artificial interference scheme
is proposed. Specifically, the destination $D$ sends a signal to interfere with reception at the
relay, while the source is transmitting to the relay. Then,
the relay sends the corrupted source signal to the destination, where it
can be recovered since the destination has full information about
the interfering signal. This two phase transmission is similar to
 physical layer network coding as described in
\cite{mobicom:06}. The data transmission process is detailed as
follows.

%Without loss a generality, We consider the capacity achieving nested polar code scheme as in [ref].
Before transmitting the signal, the source needs to encode its
information with a secure encoding scheme, such as those described
in \cite{wyner_secrecy:75}, \cite{ choo_nested_lattice:11} and
\cite{ Mahdavifar_secure_polar:11}. Essentially, all the schemes can
be described as follows. First, the length-$L_1$ binary confidential
message $v$ is combined with a random sequence and mapped to a new
binary source message $w$ of length $L_2 \geq L_1$.
%The random
%sequence is unknown to both $R$ and $D$, so that the relay can
%obtain no information about the confidential message $v$ (this
%is a one-to-one mapping and the mapping rule on how to map a given sequence $v$ and a given random sequence to a new sequence $w$ is known to all
%the nodes).
This is a one-to-one mapping, the rule for which is known to all nodes, although the random sequence itself is not known to either $R$ or $D$.
The sequence $w$ is then channel encoded to a length-$N$
message $x_s$ to combat the channel degradation at the destination.

After secure encoding at the source, the data transmission will
operate in two phases as shown in Fig. \ref{system_model}.
Specifically, the first phase is a multiple access transmission,
in which the source $S$ transmits $x_s$ with power $P_s$ to the relay,
and at the same time, $D$ transmits a Gaussian
interference signal $x_d$ with power $P_d$. Hence the relay receives a
superimposed signal
\begin{align} \label{eq:relay_signal}
 y_r=h_1 x_s+h_2 x_d+n_r,
\end{align}
where $h_{i}\sim{\cal CN}(0, \varepsilon_i)$ denotes the
instantaneous channel coefficient of the $i$-th hop ($i=1,2$) in the
multiple access phase, and $n_r \sim{\cal CN}(0, \sigma^2)$ is
additive white Gaussian noise at the relay.

Recall that the relay is untrusted, and we hope to keep the
confidential information secret from the relay. Therefore, the relay
must not be able to correctly decode $w$ from $y_r$. On the other
hand, the relay is willing to forward the message to the
destination. So it will process $y_r$ to make the transmission to
$D$ more efficient. We denote the message sent from the relay by
$x_r$, with the same average power constraint $P_s$\footnote{ There
is no loss of generality with this assumption since the transmit
power can be combined into the channel coefficient in the second
phase.}.

The second phase is a point-to-point forwarding transmission, where $R$ transmits $x_r$ to the destination $D$. As a result, the received signal at $D$ can be expressed as
\begin{align}
 \label{eq:yd}
 y_d=h_2 x_r+n_d,
\end{align}
where $n_d \sim{\cal CN}(0, \sigma^2)$ is additive white Gaussian noise at
the destination and the corresponding channel coefficient is also
$h_2$ under the assumption of reciprocity.

Let $g_i=|h_i|^2$ be the instantaneous channel gain of the $i$-th
hop ($i\in \{1,2\}$). For an amplify-and-forward relay scheme,
$x_r=\tau y_r$ with the normalizing coefficient $\tau
=\sqrt{\frac{P_s}{P_s g_1+P_d g_2+\sigma^2}}$. Although the
artificial interference $x_d$ protects the confidential message from
 eavesdropping at $R$, it also degrades the performance of the useful
information at the destination since it decreases the normalizing
coefficient $\tau$ via the term $P_d$ therein. Better protection
with larger $P_d$ also consumes more power at the relay, which is
the interference power dilemma, as noted above. To solve this
dilemma, i.e., to overcome the detrimental effects of the artificial
interference and keep its beneficial effects at the same time, we
now propose a modulo-and-forward scheme as detailed in the next
section.

\section{Modulo-and-Forward Scheme}
In the amplify-and-forward scheme, a good part of the relay's power
 is used to convey the interference, which is totally useless for
the destination $D$. To counter this problem, we propose a
modulo-and-forward scheme, in which the relay processes $y_r$ with a
modulo operation before forwarding it. This scheme is detailed as
follows.

\subsection{Lattice Encoding at S}
We consider the nested lattice coding scheme of
\cite{choo_nested_lattice:11}. Let $(\Lambda,\Lambda_0,\Lambda_1)$
be properly designed nested lattices such that $\Lambda_1 \subset
\Lambda_0 \subset \Lambda$, and their associated fundamental Voronoi
regions are denoted by $\Vc_1,\Vc_0, \text{ and } \Vc$,
respectively. The Voronoi regions are selected such that the average
power of points in $\Vc_1$ is $P_s$. All points in
$(\Lambda,\Lambda_0,\Lambda_1)$ are length-$L$ vectors. We define
the codebook $\Cc=\{ \Lambda \cap \Vc_0 \}$ and there is a
one-to-one mapping between each codeword in $\Cc$ and each
confidential message $v$. Therefore, the cardinality of $\Cc$ is
$|\Cc|=2^{L_1}$. We then define another codebook $\Cc'=\{ \Lambda_0
\cap \Vc_1 \}$ and the cardinality of this codebook is
$|\Cc'|=2^{L_2-L_1}$. Therefore, there is a one-to-one mapping
between any codeword in $\Cc'$ and any length-$(L_2-L_1)$ binary
sequence.

For any message $v$, the source selects the corresponding codeword
$c_m$ in $\Cc$, and then it generates a random bit sequence of
length $L_2-L_1$, which is mapped to one codeword $b_m$ in $\Cc'$.
Before transmission, the source $S$ calculates the nested lattice
codeword as $x_s=(c_m+b_m+u) \ \mod \ \Lambda_1$, equal to $(a_m +u)
\ \mod \ \Lambda_1$, where $u$ is a dither vector uniformly
distributed over $\Vc_1$ and $a_m=(c_m+b_m) \mod \Lambda_1$.
Obviously, the average power of $x_s$ is still $P_s$.

\subsection{Modulo Operation at $R$}
After the first phase transmission, the relay $R$ receives
$y_r=h_1x_s+h_2x_d+n_r$. Instead of forwarding $y_r$ directly as in
the AF scheme, the relay scales the received signal\footnote{With
equalization at the relay, our modulo-and-forward scheme can be
applied to both the in-phase and quadrature phase signals. For
simplicity, we focus only on the in-phase signal processing here.}
and reduces it modulo $\Lambda_1$, i.e.,
\begin{align}
 x_r&= [\beta \frac{1}{h_1}y_r + u_1 ] \ \mbox{mod} \ \Lambda_1 \nonumber \\
 &=[\beta (x_s+h_2/h_1 x_d+n_r/h_1) + u_1] \ \mbox{mod} \ \Lambda_1,
\end{align}
where $u_1$ is a random dither vector uniformly distributed over
$\Vc_1$ and is known by all the nodes, and $\beta$ is chosen as $
\frac{P_s}{P_s+\sigma^2/g_1}$ to minimize interference at the
destination as explained later.

According to the lemma in \cite{EZ:04}, $x_r$ is also uniformly
distributed over $\Vc_1$ and its average power is $P_s$. As it is
assumed that the relay has the same transmit power constraint as
$P_s$, it can directly forward the resulting signal $x_r$ to the
destination $D$.

\subsection{Lattice Decoding at $D$}
After receiving the signal $y_d$ of (\ref{eq:yd}), the destination
$D$ exploits inflated lattice decoding \cite{EZ:04}. Specifically,
$D$ multiplies the received signal by a coefficient $\alpha$, and
then cancels both the self-interference $x_d$ and
 the dither vector $u, u_1$ as
\begin{align}
 y=\Big(\frac{\alpha}{h_2}y_d-\beta \frac{h_2}{h_1}x_d-u -  u_1\Big) \ \mbox{mod} \ \Lambda_1,
\end{align}
where $\alpha$ and $\beta$ are scaling factors to be selected to
minimize the power of the residual interference plus noise. By
substituting (2) and (3) into (4) and applying
processing similar to that in \cite{EZ:04} and
\cite{Ozgur_lattice_capacity_2013}, we can write $y$ as
\begin{align} \label{eq:post_modulo_D}
 y &= \Big( \alpha (x_r+\frac{n_d}{h_2})-\beta \frac{h_2}{h_1}x_d-u - u_1\Big) \ \mbox{mod} \ \Lambda_1 \nonumber \\
 &= \Big( x_r + (\alpha-1)x_r+ \alpha \frac{n_d}{h_2} -\beta \frac{h_2}{h_1}x_d-u - u_1\Big) \ \mbox{mod} \ \Lambda_1 \nonumber \\
 &= \Big( \beta x_s+ \beta \frac{n_r}{h_1} + (\alpha-1)x_r+ \alpha \frac{n_d}{h_2} -u \Big) \ \mbox{mod} \ \Lambda_1 \nonumber \\
 &= \Big( x_s+(\beta-1) x_s+ \beta \frac{n_r}{h_1} + (\alpha-1)x_r+ \alpha \frac{n_d}{h_2} -u \Big) \ \mbox{mod} \ \Lambda_1 \nonumber \\
 &=\Big(a_m+(\alpha-1)x_r+(\beta-1)x_s+\beta\frac{n_r}{h_1}
 +\alpha\frac{n_d}{h_2}\Big) \ \mbox{mod} \
 \Lambda_1,
\end{align}
where the residual interference plus noise becomes
$\big((\alpha-1)x_r+(\beta-1)x_s+\beta\frac{n_r}{h_1}
 +\alpha\frac{n_d}{h_2}
\big) $, with an upper bound on the variance of
$(1-\alpha)^2P_s+(1-\beta)^2P_s+\alpha^2\sigma^2/g_2+\beta^2\sigma^2/g_1
$. We then select $\alpha$ and $\beta$ to minimize this upper bound
on the variance\footnote{$x_r$, $x_s$ and $a_m$ are independent of
each other with the random dither vector $u$ and $u_1$ as in
\cite{EZ:04}.}. It is easy to see that the optimal values of the
scaling factors are $\alpha= \frac{P_s}{P_s+\sigma^2/g_2}$ and
$\beta = \frac{P_s}{P_s+\sigma^2/g_1}$. As a result, the equivalent
noise variance of the post-modulo signal at $D$ becomes
\begin{align}
 \label{eq:sigmae}
 \sigma_e^2 = \min \Big\{ P_s, \frac{P_s\sigma^2}{g_1P_s+\sigma^2}+
 \frac{P_s\sigma^2}{g_2P_s+\sigma^2}\Big\},
\end{align} and we ignore the
case of $\sigma_e^2=P_s$ hereafter for simplicity.

 Then, the decoder at the destination can use Euclidean
lattice decoding to decode $a_m$ as
\begin{align}
 \hat{a}_m = \Qc_{\Vc} (y) \ \mbox{mod} \ \Lambda_1 ,
\end{align}
where $\Qc_{\Vc} (x)$ is the nearest neighbor quantizer defined as
$\Qc_{\Vc} (x) = \arg \min_{a \in \Lambda} ||x-a|| $. $\hat{a}_m$
can then be mapped to an estimate of $w$ directly (note that $w$ includes both information of confidential message $v$ and the random generated information.).

As proved in \cite{EZ:04}, the error probability $\Pr(a_m \ne
\hat{a}_m)$ goes to zero as long as the total transmission rate
$R_d=L_2/L$ is less than the direct channel capacity $C_d =
\frac{1}{2}\log(P_s/\sigma_e^2)$. We use
$C_r=\frac{1}{2}\log(1+g_1P_s/(g_2P_d+\sigma^2))$ to denote the
channel capacity for the untrusted relay (with interference). As
proved in \cite{choo_nested_lattice:11}, this nested scheme can
guarantee that the untrusted relay obtains no information about the
confidential message as long as the confidential rate $R_s=L_1/L$ is
less than the secrecy capacity $[C_d-C_r]^+$, where $[x]^+=\max (0,
x)$.

By decreasing the power of the effective signal $y_r/h_1$ from
$P_s+\frac{g_2}{g_1} P_d+\frac{\sigma^2}{g_1}$ to $P_s$, the MF
scheme can substantially improve the performance of the untrusted
relay channel. The performance of MF is analyzed in the next two
sections.

%
%\section{Performance metric of secure communication}
%In this section, we provide the performance metric of secure
%communication under different condition of CSI at the transmitter.
%In particular, three cases including full CSIT, partial CSIT, and no
%CSIT are considered.
%
%
%\subsection{Secrecy Capacity with Full CSIT}
%
%
%\subsection{secrecy Outage with Partial CSIT}
%
%In general, the wire-tapper is in a passive mode and the its
%instantaneous channel information is difficult to be known by the
%transmitter (statistical information could be known). On the other
%side, the main channel can be assumed to be known since the
%destination can help to feed back the channel information to the
%transmitter. In this case, the transmitter can set the mixed message
%rate $R_d$ equal to the main channel capacity, i.e., $R_d=C_d$ to
%obtain the best performance. However, the confidential message rate,
%$R_s\leq R_d$, can only be set in an artificial way (according to
%the statistical information of the wiretapper channel). Then, there
%is always a possibility that $R_s > [C_d-C_r]^+$, which means that
%some confidential information is leaked to the wiretapper. This
%event is defined as the secrecy outage as in
%\cite{barros_secure_outage:06, bloch_security:08}. In this case, the
%outage probability with given secrecy rate can be used to measure
%the system performance.
%
%\subsection{Outage-Leakage Performance with no CSIT}

\section{Secrecy Capacity Analysis with Full CSIT}
In this section, we analyze the capacity performance on the
untrusted relay channel, in terms of the secrecy capacity and the
generalized secure degrees of freedom, under the assumption of full
CSIT. This section consists of four parts: Part A presents the
definition of secrecy rate and generalized secure degrees of
freedom; Part B presents upper bounds for the secrecy rate and
G-SDoF for any forwarding protocol; Part C calculates the achievable
secrecy rate and G-SDoF of the MF scheme; and finally Part D
provides a comparison with the AF and DF forwarding schemes.

\subsection{Secrecy Rate and G-SDoF}
When the channel varies slowly and the channel information can be
fed back to the transmitter, all the channel information can be
pre-known by the transmitter. In this case, the transmitter can
carefully select the confidential message rate $R_s$ and the mixed
message rate $R_d$ respectively, such that the confidential message
is correctly and securely transmitted to the destination at the maximal
rate. In this case, the achievable secrecy capacity region has been
derived in \cite{wyner_secrecy:75} and is given by
\begin{align} \label{eq:secure_condition_simple}
R_s \leq R_d\leq C_d \ \ \ 0\leq R_s \leq [C_d-C_r]^+,
\end{align}
where $C_d$ and $C_r$ are the channel capacities from the source to
the destination (via the relay) and the channel capacity from source to
the relay, respectively. In this case, therefore, the secrecy
capacity is the most important factor indicating the system
performance.
%${\it{secrecy Outage}}

In the high SNR region, degrees of freedom is a good metric to
analyze the system rate performance. Analogous to the generalized
degrees of freedom definition in \cite{jafar_diversity:10}, we
further define the generalized secure degrees of freedom as
\begin{align} \label{eq:gsdof_definition}
SD(\rho) = \mathop { \lim \sup}_{SNR \to \infty }
\frac{R_s(SNR,\rho)}{\log(SNR)},
\end{align}
where $SNR$ is defined as $P_s/\sigma^2$ and
$\rho=\log(INR)/\log(SNR)$, with $INR=P_d/\sigma^2$ being the
interference-to-noise ratio.

\subsection{Upper Bound on Secrecy Rate and G-SDoF}
We firstly present an upper bound on the secrecy rate for any
possible forwarding protocol and processing at the relay node. As
proved in \cite{ wyner_secrecy:75} and \cite{leung_secrecy:78},
the secrecy rate for a wiretap channel is upper bounded by
$[C_d-C_r]^+$. In our system, the destination channel capacity $C_d$
is upper bounded by the minimum capacity of the two-hop channel
based on the cut-set bound, which is $C_d \leq
\frac{1}{2}\log(1+\min\{g_1, g_2\}\frac{P_s}{\sigma^2})$. On the
other hand, the capacity of the relay channel, $C_r =
\frac{1}{2}\log(1+\frac{g_1P_s}{g_2P_d+\sigma^2})$, is achievable with properly selected lattices.
 As a result, an upper bound on the secrecy rate for this two-hop channel is given by
\begin{align} \label{eq:upper_bound}
 U &= \left[ \frac{1}{2}\log\left(1+\min\{g_1, g_2\}\frac{P_s}{\sigma^2}\right)-\frac{1}{2}\log(1+\frac{g_1P_s}{g_2P_d+\sigma^2}) \right]^+ .
\end{align}

We secondly present an upper bound for the G-SDoF. Substituting the
upper bound in \eqref{eq:upper_bound} to the definition in
\eqref{eq:gsdof_definition}, we can easily obtain an upper bound on
the G-SDoF as
\begin{align} \label{eq:gsdof_upperbound}
SD_{u}(\rho) &=\mathop {\lim \sup}_{SNR \to \infty } \frac{\frac{1}{2}\log(1+\min\{g_1, g_2\}\frac{P_s}{\sigma^2})-\frac{1}{2}\log(1+\frac{g_1P_s}{g_2P_d+\sigma^2})}{\log(SNR)} \nonumber \\
 &=\frac{1}{2}\mathop{\lim \sup}_{SNR \to \infty } \frac{\log\Big(SNR \min(g_1, g_2)\Big) - \log\Big(1+\frac{g_1 SNR }{g_2 INR }\Big)}{\log(SNR)} \nonumber \\
 &=\frac{1}{2}\mathop{\lim\sup}_{SNR \to \infty }\Bigg[ 1-\frac{ \log\Big(1+ \frac{ g_1}{ g_2}SNR^{1-\rho}\Big)}{\log(SNR)}\Bigg] \nonumber \\
 &= \left\{
 \begin{array}{ll}
 \displaystyle
 \rho/2 ,\hspace{3mm} \mbox{If}\ 0 \leq \rho < 1
 \vspace{3mm}\\
 \displaystyle
 1/2 , \hspace{3mm} \mbox{If}\ 1 \leq \rho
 \end{array}
 \right..
\end{align}
The G-SDoF must be no more than the generalized degrees of
freedom without a security constraint, which is at most 1/2 in
our two-hop single antenna system due to the two-phase transmission.
And this best secure DoF may be achieved when the transmission power
of the destination is no less than that of the source, as shown in
\eqref{eq:gsdof_upperbound}.

\subsection{Achievable Secrecy Rate and G-SDoF with MF}
We now calculate the achievable secrecy rate and G-SDoF for the
modulo-and-forward scheme.

\subsubsection{Achievable Secrecy Rate}

%With reference to \eqref{eq:post_modulo_D}, the
%resultant SNR at the destination is
%\begin{align}
% \mbox{SNR}_d=\frac{P_s}{\sigma^2}\frac{g_1 g_2}{g_1+g_2}.
%\end{align}
%With the good lattice code, the achievable rate of the main link is
%\begin{align}
% R_d=\frac{1}{2}\log\Big(1+\frac{P_s}{\sigma^2}\frac{g_1
% g_2}{g_1+g_2}\Big).
%\end{align}

With reference to \eqref{eq:sigmae},
 {a} good nested lattice code can achieve the following rate to the destination:
\begin{align}
 R_d&=\frac{1}{2}\log\Big(\frac{P_s}{\sigma_e^2}\Big) \nonumber \\
 &=\frac{1}{2}\log\Big(\frac{1}{\frac{\sigma^2}{g_1P_s+\sigma^2} + \frac{\sigma^2}{g_2P_s+\sigma^2} }\Big) \nonumber \\
 &\geq \frac{1}{2}\log\Big( \frac{1}{2} + \frac{P_s}{\sigma^2}\frac{g_1g_2}{g_1+g_2}\Big).
\end{align}

%
%\begin{remark}
%\textcolor[rgb]{1.00,0.00,0.00}{Note that the above $R_d$ is
%even better than the rate of a two-hop pure relay channel with the
%traditional amplify-and-forward scheme at the relay without jamming
%signal from $D$}. In other words, the interference transmission of
%the destination did not degrade the performance of the main link,
%providing an extra benefit of security.
%\end{remark}

From \eqref{eq:relay_signal}, we can compute the maximum information
obtained by the relay. With the assumption that the relay knows all
the nested lattice codebook information and the channel state
information to detect the combined information $w$, the maximal
information rate is
\begin{align}
 R_r=\frac{1}{2}\log\Big(1+\frac{P_s g_1}{P_d
 g_2+\sigma^2}\Big).
\end{align}

Then, the achievable secrecy rate of the proposed modulo-and-forward
scheme is given by
\begin{align} \label{eq:achievable_rate}
 R_s & \geq \frac{1}{2}\left[\log\Big(\frac{1}{2}+\frac{P_s}{\sigma^2}\frac{g_1
 g_2}{g_1+g_2}\Big) - \log\Big(1+\frac{P_s g_1}{P_d
 g_2+\sigma^2}\Big) \right]^+ .
\end{align}

From \eqref{eq:achievable_rate}, we can see that the artificial noise power
$P_d$ only helps to improve the secrecy rate with almost no detrimental
effect on the transmission from $S$ to $D$ in our modulo-and-forward
scheme. When $P_d$ goes to infinity, $R_s$ can approach the upper
bound of $\frac{1}{2}\log\Big(1/2+\frac{P_s}{\sigma^2}\frac{g_1
g_2}{g_1+g_2}\Big)$. Therefore, the MF scheme solves the
interference power dilemma of the AF scheme.

We now calculate the gap between the upper bound and the achievable rate in the non-trivial regime, i.e., $U > 0$ and $R_s > 0$. Without loss of generality, we assume that $g_1 \leq g_2$. Then, we obtain the gap as
%\begin{align}
% U-R_s &= \frac{1}{2}\log(1/2+g_1\frac{P_s}{\sigma^2}) - \frac{1}{2}\log\Big(1+\frac{P_s}{\sigma^2}\frac{g_1
% g_2}{g_1+g_2}\Big) \nonumber \\
% &=\frac{1}{2}\log\Big( 1+ \frac{g_1}{g_2} \frac{g_1P_s/\sigma^2}{1+\frac{g_1}{g_2} +g_1P_s/\sigma^2} \Big) \nonumber \\
% &\leq \frac{1}{2}\log\Big( 1+ \frac{g_1}{g_2} \Big) \nonumber \\
% &\leq 1/2,
%\end{align}
\begin{align}
 U-R_s %\frac{1}{2}\log(1+g_1\frac{P_s}{\sigma^2}) - \frac{1}{2}\log\Big(1+\frac{P_s}{\sigma^2}\frac{g_1
% g_2}{g_1+g_2}\Big) \nonumber \\
 &\leq\frac{1}{2}\log\Big( 1+ \frac{g_1+g_2+2g_1^2P_s/\sigma^2}{g_1+g_2+2g_1g_2P_s/\sigma^2} \Big)
 \leq 1/2,
\end{align}
which goes to zero as ${g_1}/{g_2}$ goes to zero in the high SNR
region. In other words, when the ratio between the two-hop channel
gains, $g_1$ and $g_2$, becomes very large or small, the achievable
rate of the modulo-and-forward scheme approaches the upper bound for
small noise variance. In summary, we have the following theorem,

\begin{theorem}
The modulo-and-forward scheme achieves a secrecy rate $R_s$ of
\eqref{eq:achievable_rate},
%\begin{align}
%R_s = \frac{1}{2}\left[\log\Big(1+\frac{P_s}{\sigma^2}\frac{g_1
% g_2}{g_1+g_2}\Big) - \log\Big(1+\frac{P_s g_1}{P_d
% g_2+\sigma^2}\Big) \right]^+,
%\end{align}
which is within one-half bit of the secrecy capacity for all channel
realizations{\footnote{ The above analysis ignores the case that
$U>0$ and $R_s<0$. In fact, it is easy to verify that $U < 1/2$ when
$R_s<0$. Therefore, this theorem is true for all channel
realizations.}}.
\end{theorem}

\subsubsection{Achievable G-SDoF}
Now we characterize the G-SDoF of the MF scheme.
Substituting \eqref{eq:achievable_rate} into the definition of G-SDoF in \eqref{eq:gsdof_definition}, we have
\begin{align}
SD_{mf}(\rho) &\geq \mathop{\lim \sup}_{SNR \to \infty }
\frac{\frac{1}{2}\log\Big(1/2+\frac{P_s}{\sigma^2}\frac{g_1
 g_2}{g_1+g_2}\Big) - \frac{1}{2}\log\Big(1+\frac{P_s g_1}{P_d
 g_2+\sigma^2}\Big)}{\log(SNR)} \nonumber \\
 &=\frac{1}{2}\mathop{\lim \sup}_{SNR \to \infty } \frac{\log\Big(SNR\frac{g_1
 g_2}{g_1+g_2}\Big) - \log\Big(1+\frac{SNR g_1}{INR g_2}\Big)}{\log(SNR)} \nonumber \\
 &=\frac{1}{2}\mathop{\lim \sup}_{SNR \to \infty } 1-\frac{ \log\Big(1+ \frac{ g_1}{ g_2}SNR^{1-\rho}\Big)}{\log(SNR)} \nonumber \\
 &= \left\{
 \begin{array}{ll}
 \displaystyle
 \rho/2 ,\hspace{3mm} \mbox{If}\ 0 \leq \rho < 1
 \vspace{3mm}\\
 \displaystyle
 1/2 , \hspace{3mm} \mbox{If}\ 1 \leq \rho
 \end{array}
 \right..
\end{align}
As an upper bound on G-SDoF, $SD_{u}(\rho)\geq SD_{mf}(\rho)$ holds.
On the other hand, $SD_{mf}(\rho) \geq SD_{u}(\rho)$ is obtained
from the equation above. Hence we can conclude that $SD_{mf}(\rho) = SD_{u}(\rho)$ exactly. Thus, we have
the following theorem,
\begin{theorem}
The modulo-and-forward scheme {achieves} the full generalized secrecy degrees of freedom for the untrusted relay channel.
\end{theorem}

\subsection{Comparison with AF and DF Schemes}

This subsection compares the MF scheme with the AF scheme in terms
of secrecy rate and the generalized secure degrees of freedom.

\subsubsection{Secrecy Rate Comparison with AF}

For the AF scheme, the relay node will amplify the received signal
in \eqref{eq:relay_signal} with a coefficient $\tau =
\sqrt{P_s/(g_1P_s+g_2P_d+\sigma^2)}$ before sending it to the
destination $D$. As a result, the destination receives the signal
\begin{align}
y_d ' &= \tau h_2 y_r+n_d \nonumber \\
&=\tau h_1h_2 x_s + \tau h_2^2 x_d + \tau h_2 n_r + n_d.
\end{align}
Since the destination has full knowledge of $x_d$ and the channel
coefficients, it can completely remove the self-interference term
$\tau h_2^2 x_d$, and the {resulting} SNR of the target signal
becomes
\begin{align}
SNR_{AF}=\frac{P_s^2 g_1 g_2}{\sigma^2
[P_sg_1+P_sg_2+P_dg_2+\sigma^2 ]}.
\end{align}
Then, we can calculate the secrecy rate of the traditional AF scheme as \cite{AFRelayCapacity}
\begin{align} \label{eq:af_rate}
R_{af}= \left[ \frac{1}{2}\log \big(1+\frac{P_s^2 g_1 g_2}{\sigma^2
[P_sg_1+P_sg_2+P_dg_2+\sigma^2 ]}\big) - \frac{1}{2}\log \big(1+
\frac{P_s g_1}{P_d
 g_2+\sigma^2} \big) \right]^+.
\end{align}

By comparing $R_{AF}$ with $R_s$ in \eqref{eq:achievable_rate}, it
is easy to verify that the secrecy rate of the MF scheme, $R_s$, is
always larger than that of the traditional AF scheme. Specifically,
consider the case in which $\sigma^2 \to 0, P_d \to \infty$, with
$P_d*\sigma^2$ and other parameters constant. Then $R_{AF}$ in
\eqref{eq:af_rate} is a constant while the rate $R_s$ in
\eqref{eq:achievable_rate} increases without bound. In other words,
the improvement of the MF scheme over the AF scheme can be
arbitrarily large.

%So, we have the following corollary.
%\begin{corollary}
% The secrecy rate of our modulo-and-forward scheme is always larger than that of the traditional amplify-and-forward scheme as long as it is not zero, and the gap can be infinity when the noise variance goes to zero and the power of destination goes to infinity.
%\end{corollary}

\begin{figure}
\begin{center}
\includegraphics[width=5.0in]{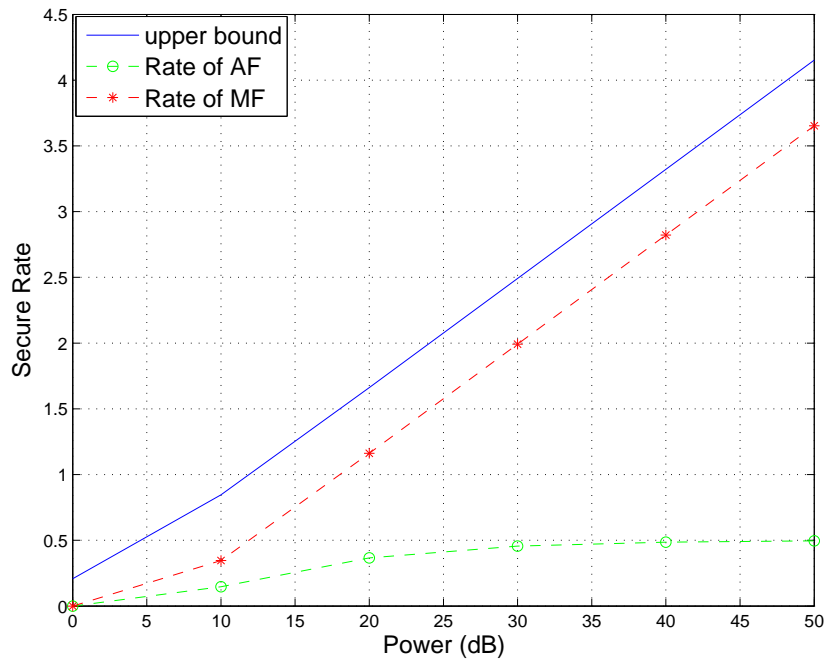}
\caption{Secrecy rate versus $P_d$, where $P_s=\sqrt{P_d}$ and
$\sigma^2=1$. } \label{capacity_power_label}
\end{center}
\end{figure}

In Fig. \ref{capacity_power_label}, we plot the achievable secrecy
rate of the MF and AF schemes, as well as the upper bound in
\eqref{eq:upper_bound} versus transmission power $P_d$. In
particular, we set $g_1=g_2=1$, $P_s=\sqrt{P_d}$ and $\sigma=1$. The
figure verifies that the rate of the MF scheme and the upper
bound increase with the transmission power, and their gap is always
less than $1/2$. On the other hand, the rate of the AF scheme
approaches a constant of $1/2$.

\subsubsection{G-SDoF Comparison with AF}

Since the rate gap between the MF and the AF schemes can be
arbitrarily large in the high SNR region, their G-SDoF should be
different. Here, we calculate the G-SDoF of the AF scheme to make a
comparison. According to the definition in
\eqref{eq:gsdof_definition}, we can calculate the G-SDoF of the AF
scheme as follows:
\begin{align} \label{eq:gsdof_af}
SD_{af}(\rho) &= \mathop{\lim \sup}_{SNR \to \infty }
\frac{\frac{1}{2}\log \big(1+\frac{P_s^2 g_1 g_2}{\sigma^2
[P_sg_1+P_sg_2+P_dg_2+\sigma^2 ]}\big) - \frac{1}{2}\log \big(1+
\frac{P_s g_1}{P_d
 g_2+\sigma^2} \big)}{\log(SNR)} \nonumber \\
 &=\frac{1}{2}\mathop{\lim \sup}_{SNR \to \infty } \frac{\log\Big(1+SNR\frac{SNR g_1
 g_2}{SNR(g_1+g_2)+INRg_2}\Big) - \log\Big(1+\frac{SNR g_1}{INR g_2}\Big)}{\log(SNR)} \nonumber \\
 &=\frac{1}{2}\mathop{\lim \sup}_{SNR \to \infty } \frac{\log\Big(1+SNR^{2-\rho}\frac{g_1
 g_2}{SNR^{1-\rho}(g_1+g_2)+g_2}\Big) - \log\Big(1+\frac{g_1}{g_2}SNR^{1-\rho}\Big)}{\log(SNR)} \nonumber \\
 &= \left\{
 \begin{array}{ll}
 \displaystyle
 \rho/2 ,\hspace{3mm} \mbox{If}\ 0 \leq \rho < 1
 \vspace{3mm}\\
 \displaystyle
 1-\rho/2 , \hspace{3mm} \mbox{If}\ 1 \leq \rho < 2
 \vspace{3mm}\\
 \displaystyle
 0 , \hspace{3mm} \mbox{If}\ 2 \leq \rho
 \end{array}
 \right..
\end{align}

\begin{figure}
\begin{center}
\includegraphics[width=5.0in]{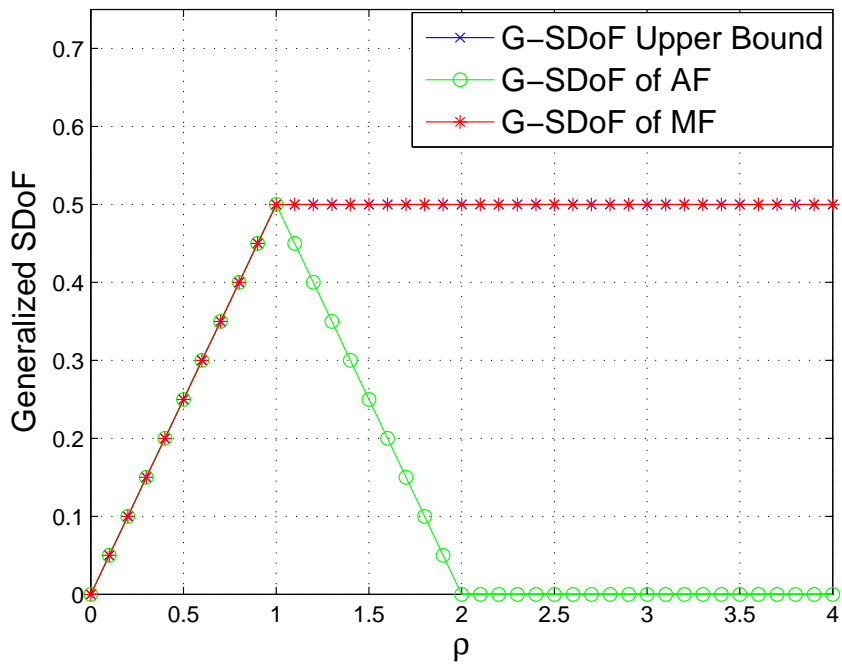}
\caption{Generalized secure degrees of freedom versus $\rho$ .}
\label{gsdof_label}
\end{center}
\end{figure}
 We plot the G-SDoF of AF and MF in Fig. \ref{gsdof_label} for an intuitive comparison.
 With reference to \eqref{eq:af_rate}, its first term,
 $\frac{1}{2}\log \big(1+\frac{P_s^2 g_1 g_2}{\sigma^2 [P_sg_1+P_sg_2+P_dg_2+\sigma^2 ]}\big)$,
 is a decreasing function of $P_d$ while the second term,
 $- \frac{1}{2}\log \big(1+ \frac{P_s g_1}{P_d g_2+\sigma^2} \big)$,
 is an increasing function of $P_d$. Hence, the destination needs to carefully select an optimal value of $P_d$ to maximize the rate $R_{af}$ if all the channel information is also available at the destination (this is the interference power dilemma). In the high SNR regime, $P_d$ is easier to calculate with reference to \eqref{eq:gsdof_af}, i.e., $P_d$ and $P_s$ should have the same order
 ($\rho=1$).

\subsubsection{Comparison with DF}

In the AF and MF schemes, only symbol level time synchronization is
required at the multiple access phase. Currently, there is no
capacity approaching DF scheme under this assumption. In this part,
we compare the MF scheme with the lattice based DF scheme in
\cite{he_lattice:13}, where perfect phase, amplitude and time
synchronization between the interfering signal and the source signal
are assumed.

With reference to \cite{he_lattice:13}, setting ${h_1=h_2 }$, and
$P_s=P_d$, the achievable rate of lattice DF scheme is
$$R_{df}=\frac{1}{2} \log\left(\frac{1}{2}+\frac{P_sg_1}{\sigma^2}\right)-1.$$

With the same channel coefficients, the MF scheme can also achieve a
larger secure rate even without such a strict synchronization
requirement. The secrecy rate improvement can be calculated as
\begin{align} \label{eq:df_rate_difference}
R_s-R_{df} = \left\{
 \begin{array}{ll}
 \displaystyle
 0 \hspace{3mm} \mbox{If}\ t\leq1
 \vspace{3mm}\\
 \displaystyle
 \frac{1}{2}\log(1+t)-\frac{1}{2}
 \hspace{3mm} \mbox{If}\ 1 < t \leq 3/2
 \vspace{3mm}\\
 \displaystyle
 \frac{1}{2}\log\Big(2+\frac{2}{1+2t}\Big)
 \hspace{3mm} \mbox{If}\ t > 3/2
 \end{array}
 \right.,
\end{align}
where $t=P_sg_1/\sigma^2$ is the receiver side SNR. From \eqref{eq:df_rate_difference}, we
can see that the gap will always {be} larger than 0 for the non-trivial case.
 For the low SNR region, the improvement of the MF scheme could be
significant since $R_s/R_{df}$ can goes to infinity\footnote{The
achievable rate in \cite{he_lattice:13} may be further improved with
a better bounding technique. However, the MF scheme should outperform
it at least in the low SNR region.}.

\begin{figure}
\begin{center}
\includegraphics[width=5.0in]{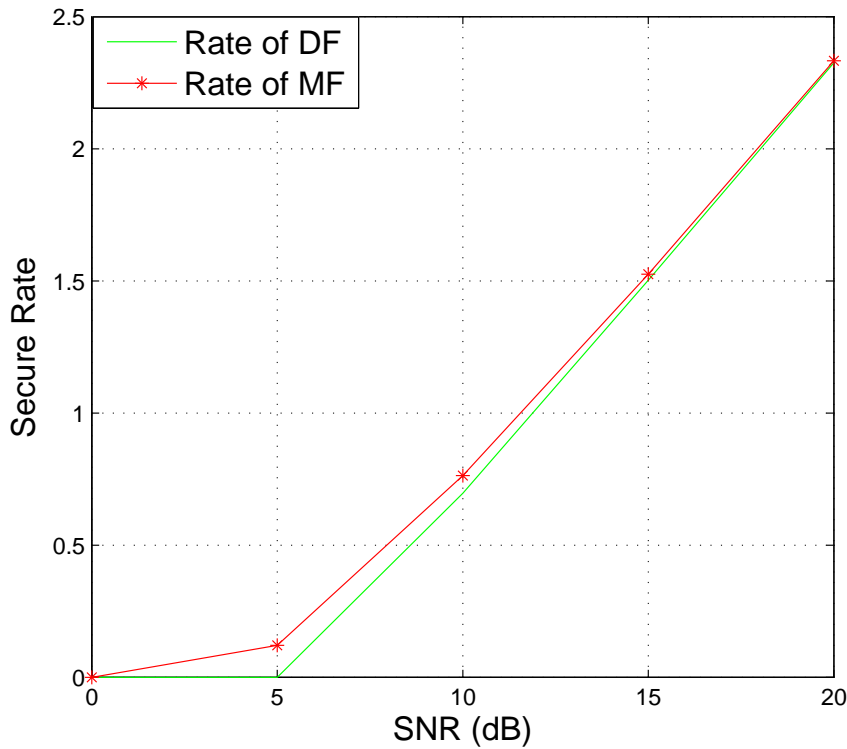}
\caption{Secure capacity versus $SNR=P_sg_1/\sigma^2$ .} \label{capacity_df_label}
\end{center}
\end{figure}

\section{Outage Performance without CSIT}
In this section, we analyze the connection outage and secrecy outage
performance of the MF scheme under the assumption of only receiver
side channel state information. Specifically, this section consists
of four parts: Part A presents the formulation of connection and
secrecy outage probabilities, as well as the generalized secure
diversity; Parts B and C calculate a lower bound and the achievable
probability for the outage probabilities, as well as the secure
diversity; and Part D provides a comparison with the AF scheme.

\subsection{Outage Probabilities and Diversity}
The metrics of secrecy capacity and secrecy outage have been widely
investigated to measure the performance of various schemes. However,
they are not appropriate when the transmitter has no channel state
information. In many cases, such as the fast fading case
\cite{outdated_csi_2009}, even the state information of the
source-destination channel is difficult to obtain at the
transmitter. Of course, with an untrusted channel state information
provider, such as the relay in our system model, other issues arise.

%{ One possible reason is the fast fading channels in
%wireless communication systems. The channel state usually
%significantly changes when the CSI is calculated based on the last message and feedback
%from the receiver side \cite{outdated_csi_2009}. Another possible reason is the
%untrusted nature of the CSI provider, such as the relay in our system
%model. Specifically, the relay in our model only needs to provide
%$h_2/h_1$ for the detection at the receiver, and it may provide inaccurate individual channel information of the four channel coefficients for the transmitter
%to calculate the channel capacities \footnote{without assumption of
%reciprocity, there are generally four channels for both directions
%and both hops.}. }

Without any CSIT, both the direct channel rate $R_d$ and the secrecy
rate $R_s$ must be determined before transmission. Thus, there is
always the possibility of unsuccessful transmissions. There are of
two types of unsuccessful transmissions to consider: (1) secrecy
outage, in which the eavesdropper obtains some information about the
confidential message; and (2) connection outage, in which the
destination fails to detect the mixed and confidential messages. We
use the C-S outage (connection outage and secrecy outage)
probability to measure the system secure performance
\cite{tang_connection_outage_2009}. We formulate the C-S performance
as follows.

\subsubsection{Outage Probabilities Formulation}
To discuss the outage probabilities, we define three events: (1)
$E_A$: secrecy outage occurs at the relay and no connection outage
{occurs} at the destination; (2) $E_B$: connection outage occurs at
the destination and no secrecy outage occurs at the relay; and (3)
$E_C$: both connection outage and secrecy outage occur.

A connection outage event occurs when the receiver cannot correctly decode
the received message. The connection outage probability can be written as
\begin{align} \label{eq:connection_outage_def}
\po=\Pr(E_B \cup E_C) = \Pr( C_d < R_d).
\end{align}

A secrecy outage event occurs when the eavesdropper can obtain some
information about the confidential message. This event happens when
the total information the eavesdropper can retrieve from the received
message is more than the entropy of the random sequence. So, the
secrecy outage probability can be written as
\begin{align} \label{eq:secrecy_outage_def}
\pl=\Pr(E_A \cup E_C)=\Pr( R_d-R_s < C_r)=\Pr(R_s > R_d-C_r).
\end{align}

%{ need proof? the form of leakage is a little different from the outage since it is the probability that the channel capacity is bigger than given threshold}

Both $\po$ and $\pl$ are of interest in this paper since the
connection outage and secrecy outage may be of different severities
in various scenarios. For a given $R_s$, there is a tradeoff between
the connection outage probability and secrecy outage probability,
since a large $R_d$ will decrease the secrecy outage probability
while increasing the connection outage probability, and vice versa.

\subsubsection{Definition of G-SDG}

In {the} high SNR region, diversity order is a simple metric to
characterize the system performance. We now define the generalized
secure diversity gain. Similar to the generalized diversity
gain defined in \cite{lo_generalized_diversity:09}, the G-SDG is
defined as
\begin{align} \label{eq:gsdg_definition}
DG(\rho) = \mathop{\lim \sup}_{SNR \to \infty } \frac{-\log
p_{t}(SNR,\rho)}{\log(SNR)}
\end{align}
where $SNR = P_s/\sigma^2$, an exact expression for $\rho=\log(INR)/\log(SNR)$ are
defined the same as {for} the G-SDoF in \eqref{eq:gsdof_definition},
and $p_{t}$ is the total outage probability
$p_{t}=\Pr(E_A)+\Pr(E_B)+\Pr(E_C)$.

\subsection{Bounds on Outage Probabilities and Diversity Orders}
This section develops a lower bound on the connection outage
probability, an expression for the secrecy outage probability, and an upper bound
on the G-SDG, for any possible processing at the relay node.

\subsubsection{Lower Bound on Connection Outage}
With reference to the definition in
\eqref{eq:connection_outage_def}, a lower bound on the connection
outage probability can be obtained when the direct channel capacity $C_d$ is
replaced with its upper bound $\frac{1}{2}\log(1+\min\{g_1,g_2\}
P_s/\sigma^2)$. Then, we have a lower bound on the connection
outage probability as
\begin{align}
 \polb&=\Pr\Big( \frac{1}{2}\log(1+\min\{g_1,g_2\} P_s/\sigma^2) < R_d\Big)
 \\
 &=\Pr\Big( \min\{g_1,g_2\} P_s/\sigma^2< \gamma_{o}\Big),
\end{align}
where $\gamma_{o}=2^{2R_d}-1$ is the SNR threshold associated with
connection outage. We further derive $\polb$ as
\begin{align}
 \polb&=1-\Pr\Big( \min\{g_1,g_2\} P_s/\sigma^2\geq
 \gamma_{o}\Big)\\
 &=1-\Pr\Big( g_1 \geq \frac{\gamma_{o}\sigma^2} {P_s}\Big) \Pr\Big( g_2 \geq \frac{\gamma_{o}\sigma^2} {P_s}\Big)\\
 &=1-e^{-(\frac{1}{\varepsilon1}+\frac{1}{\varepsilon2})\frac{\gamma_{o}\sigma^2}{P_s}}.
\end{align}
where the last equation is obtained by noting the probability density functions
$f_{g_1}(x)=\frac{1}{\varepsilon_1}e^{-\frac{x}{\varepsilon_1}}$ and
$f_{g_2}(x)=\frac{1}{\varepsilon_2}e^{-\frac{x}{\varepsilon_2}}$.

\subsubsection{Expression for the Secrecy Outage}
The secrecy outage does not depend on the processing at the relay
node. Therefore, we do not need to provide a bound, but rather can deal with the exact outage
probability. With the secrecy outage probability definition in
\eqref{eq:secrecy_outage_def}, we derive an analytical expression
for the secrecy outage probability as follows:
\begin{align}
 \pl &=\Pr\big(C_r > R_d-R_s\big) \notag
 \\
 &=\Pr\Big(\frac{P_s g_1}{P_d g_2+\sigma^2}>\gamma_s\Big),
\end{align}
where $\gamma_s=2^{2(R_d-R_s)}-1$ is the secrecy outage SNR threshold. We
further write $\pl$ as
\begin{align}
 \pl&=\Pr\big(P_s g_1>\gamma_s P_d g_2+\gamma_s \sigma^2\big) \notag
 \\&=\Pr\Big(g_1>\frac{\gamma_s P_d}{P_s}g_2+\frac{\gamma_s
 \sigma^2}{P_s}\Big)
 \\&=\int_0^\infty f_{g_2}(g_2)\Big(\int_{\frac{\gamma_s P_d}{P_s}g_2+\frac{\gamma_s
 \sigma^2}{P_s}}^\infty f_{g_1}(g_1)dg_1 \Big)dg_2. \notag
\end{align}
Applying the probability density functions
$f_{g_1}(x)=\frac{1}{\varepsilon_1}e^{-\frac{x}{\varepsilon_1}}$ and
$f_{g_2}(x)=\frac{1}{\varepsilon_2}e^{-\frac{x}{\varepsilon_2}}$ in
the above equation, we can obtain an analytical expression of $\pl$
as follows:
\begin{align} \label{eq:leakage_probability}
 \pl &=\frac{1}{\varepsilon_2}e^{-\frac{\gamma_s \sigma^2}{P_s \varepsilon_1}}
 \int_0^\infty e^{-(\frac{1}{\varepsilon_2}+\frac{\gamma_s P_d}{P_s \varepsilon_1})g_2}dg_2 \notag
 \\
 &=\frac{P_s \varepsilon_1}{P_s \varepsilon_1+P_d \varepsilon_2
 \gamma_s}e^{-\frac{\gamma_s \sigma^2}{P_s \varepsilon_1}}.
\end{align}

By applying the approximation $e^{x}=1+x$ for small values of $|x|$
again, we arrive at an asymptotic expression for the secrecy outage
probability with high transmit power $P_s$, namely
\begin{align} \label{eq:simple_leakage_probability}
 \pl &\simeq \frac{P_s \varepsilon_1}{P_s \varepsilon_1+P_d \varepsilon_2
 \gamma_s} \Big(1-{\frac{\gamma_s \sigma^2}{P_s \varepsilon_1}}\Big).
 %\notag \\
 %&= \frac{\varepsilon_1}{\varepsilon_1+\rho ' \varepsilon_2
 %\gamma_s} (1-{\frac{\gamma_s \sigma^2}{P_s \varepsilon_1}}).
\end{align}

\subsubsection{Upper Bound on G-SDG}

According to the definition, an outage event occurs when either a secrecy outage or a connection outage occurs. Therefore, we have a lower bound on the total outage probability:
\begin{align}
p_{t} \geq \max\{\polb,\pl\}.
\end{align}
With the generalized secure diversity gain defined in
\eqref{eq:gsdg_definition}, we can then obtain an upper bound on the
G-SDG as
 \begin{align} \label{eq:cs_diversity_upperbound}
 G\text{-}SDG_{up}(\rho) & = \mathop{\lim \sup}_{SNR \to \infty } \frac{-\log ( \max [ \polb(SNR,\rho), \pl(SNR,\rho)] )}{\log(SNR)} \\
 & = \mathop{\lim \sup}_{SNR \to \infty } \frac{ - \max [ \log \polb(SNR,\rho), \log \pl(SNR,\rho) ] }{\log(SNR)} \\
 &= \mathop{\lim \sup}_{SNR \to \infty } \frac{ \min [ -\log ((\frac{1}{\varepsilon_1}+\frac{1}{\varepsilon_2})\frac{\gamma_{o}}{SNR}), -\log (\frac{P_s \varepsilon_1}{P_s \varepsilon_1+P_d \varepsilon_2 \gamma_s} (1-{\frac{\gamma_s }{SNR \varepsilon_1}})
 )]}{\log(SNR)} \\
 & = \min \Big[1, \mathop{\lim \sup}_{SNR \to \infty } \frac{-\log ( \frac{SNR \varepsilon_1}{SNR \varepsilon_1+SNR^\rho \varepsilon_2 \gamma_s} ) }{\log(SNR)} \Big] \\
 & = \left\{
 \begin{array}{ll}
 \displaystyle
 0 \hspace{3mm} \mbox{If}\ \rho \leq 1
 \vspace{3mm}\\
 \displaystyle
 \rho-1 \hspace{3mm} \mbox{If}\ 1 < \rho \leq 2
 \vspace{3mm}\\
 \displaystyle
 1 \hspace{3mm} \mbox{If}\ \rho > 2
 \end{array}
 \right..
\end{align}

\subsection{Achievable Outage Probabilities and Diversity}

In this section, we first derive the probabilities $\po$ and $\pl$
achievable with the MF scheme, separately. With these probabilities,
we then obtain the achievable generalized secure diversity gain of
the MF scheme. The achievable $\pl$ is the same as the expression in
\eqref{eq:leakage_probability}.

\subsubsection{Achievable Connection Outage Probability}

The achievable connection outage probability can be written as
\begin{align}
 \po&=\Pr( C_d < R_d) \notag
 \\
 \label{eq:App:eq1}
 &=\Pr\Big[\frac{1}{2}\log_2\Big(1/2+\frac{P_s}{\sigma^2}\frac{g_1 g_2}{g_1+g_2}\Big)<R_d\Big]
 \\
 \label{eq:App:eq2}
 &=1-e^{-(\frac{1}{\varepsilon_1}+\frac{1}{\varepsilon_2})\frac{\gamma_{1}\sigma^2}{P_s}}
 \frac{2\gamma_{1}\sigma^2}{P_s \sqrt{\varepsilon_1
 \varepsilon_2}}K_1\Big( \frac{2\gamma_{1}\sigma^2}{P_s \sqrt{\varepsilon_1
 \varepsilon_2}}\Big),
\end{align}
where $\gamma_{1}=\gamma_{o}-1/2$ and $K_1(x)$ denotes the first-order modified Bessel function of
the second kind, and the derivation from \eqref{eq:App:eq1} to
\eqref{eq:App:eq2} is given in Appendix I.

By applying the approximation of $K_1(x)\simeq \frac{1}{x}$ and
$e^{x}=1+x$ for small values of $|x|$, we arrive at an asymptotic
expression for the connection outage probability with high transmit
power $P_s$ as
\begin{align} \label{eq:simple_outage_probability}
 \po \simeq
 \Big(\frac{1}{\varepsilon_1}+\frac{1}{\varepsilon_2}\Big)\frac{\gamma_{1}\sigma^2}{P_s}.
\end{align}

%With the generalized secure diversity gain defined in
%\eqref{eq:gsdg_definition}, the upper bound of G-SDG can be calculated as
%follows,
%\begin{align} \label{eq:outage_diversity_upperbound}
% G\text{-}SDG(\rho) & = - \mathop{\lim \sup}_{SNR \to \infty } \frac{-\log ( \max [ \po(SNR,\rho), \pl(SNR,\rho)] )}{\log(SNR)} \\
% & = - \mathop{\lim \sup}_{SNR \to \infty } \frac{ - \max [ \log \po(SNR,\rho), \log \pl(SNR,\rho) ] }{\log(SNR)} \\
% &= \mathop{\lim \sup}_{SNR \to \infty } \frac{ \min [ -\log ((\frac{1}{\varepsilon_1}+\frac{1}{\varepsilon_2})\frac{\gamma_{o}}{SNR}), -\log (\frac{P_s \varepsilon_1}{P_s \varepsilon_1+P_d \varepsilon_2 \gamma_s} (1-{\frac{\gamma_s }{SNR \varepsilon_1}})
% )]}{\log(SNR)} \\
% & = \min \Big[1, \mathop{\lim \sup}_{SNR \to \infty } \frac{-\log ( \frac{SNR \varepsilon_1}{SNR \varepsilon_1+SNR^\rho \varepsilon_2 \gamma_s} ) }{\log(SNR)} \Big] \\
% & = \left\{
% \begin{array}{ll}
% \displaystyle
% 0 \hspace{3mm} \mbox{If}\ \rho \leq 1
% \vspace{3mm}\\
% \displaystyle
% \rho-1 \hspace{3mm} \mbox{If}\ 1 < \rho \leq 2
% \vspace{3mm}\\
% \displaystyle
% 1 \hspace{3mm} \mbox{If}\ \rho > 2
% \end{array}
% \right..
%\end{align}

\subsubsection{Tradeoff between Achievable $\po$ and $\pl$}
With reference to \eqref{eq:leakage_probability} and
\eqref{eq:App:eq2}, we see that the two probabilities are not
independent. An increase in $\po$ ($\pl$) will lead to a decrease in
$\pl$ ($\po$), which is a tradeoff mentioned in
\cite{tang_connection_outage_2009}. With the given tradeoff, it
would be interesting to carefully design the rates $R_s$, $R_d$ and
the powers $P_s$, $P_d$ to minimize the connection outage and
secrecy outage probabilities. In the following Fig.
\ref{ol_probability_label}, we plot the tradeoff between the
connection outage and secrecy outage probabilities for a typical
setting.

In the high SNR regime with large $P_s/\sigma^2$, the tradeoff
between the two probabilities is simpler. Substituting
\eqref{eq:simple_outage_probability} into
\eqref{eq:simple_leakage_probability}, we obtain an explicit
relation between $\pl$ and $\po$ as
\begin{align}
 \frac{P_s \varepsilon_1+P_d \varepsilon_2
 \gamma_s}{P_s \varepsilon_1} \pl +{\frac{\gamma_s \varepsilon_2}{\gamma_1 (\varepsilon_1+\varepsilon_2)}}\po = 1.
 %\notag \\
 %&= \frac{\varepsilon_1}{\varepsilon_1+\rho ' \varepsilon_2
 %\gamma_s} (1-{\frac{\gamma_s \sigma^2}{P_s \varepsilon_1}}).
\end{align}

\subsubsection{Achievable G-SDG}

In this section, we first give an upper bound and a lower bound on
the total outage probability of the MF scheme. Fortunately, both
bounds lead to the same G-SDG, which is also the best diversity gain
in theory.

\textbf{Upper Bound and Lower Bound for Achievable $p_{t}$: }

It is difficult to directly calculate the total outage probability
due to the dependence between $\po$ and $\pl$. Therefore, we analyze
 an upper bound and lower bound on the total outage probability,
with the MF scheme. According to the definition, we can obtain an
upper bound:
\begin{align}
p_{t}&=\Pr(E_A)+\Pr(E_B)+\Pr(E_C) \leq \Pr(E_A \cup E_C)+\Pr(E_B \cup E_C) \\
 &=\po+\pl \leq 2\max \{\po,\pl\}.
\end{align}
On the other hand, either secrecy outage or connection outage means the outage of the transmission. Therefore, we have a lower bound:
\begin{align}
p_{t} \geq \max\{\po,\pl\}.
\end{align}
We can see that the lower bound is one-half of the upper bound.

\textbf{Achievable G-SDG:}

With the generalized secure diversity gain defined in
\eqref{eq:gsdg_definition}, the diversity gain is the same for both
the upper bound and the lower bound on $p_{t}$ since they only
differ by a constant coefficient, and it can be calculated as
follows:
\begin{align} \label{eq:achievable_ol_diversity}
 G\text{-}SDG(\rho) & = \mathop{\lim \sup}_{SNR \to \infty } \frac{-\log ( \max [ \po(SNR,\rho), \pl(SNR,\rho)] )}{\log(SNR)} \\
 & = \mathop{\lim \sup}_{SNR \to \infty } \frac{ - \max [ \log \po(SNR,\rho), \log \pl(SNR,\rho) ] }{\log(SNR)} \\
 &= \mathop{\lim \sup}_{SNR \to \infty } \frac{ \min [ -\log ((\frac{1}{\varepsilon_1}+\frac{1}{\varepsilon_2})\frac{\gamma_{1}}{SNR}), -\log (\frac{P_s \varepsilon_1}{P_s \varepsilon_1+P_d \varepsilon_2 \gamma_s} (1-{\frac{\gamma_s }{SNR \varepsilon_1}})
 )]}{\log(SNR)} \\
 & = \min \Big[1, \mathop{\lim \sup}_{SNR \to \infty } \frac{-\log ( \frac{SNR \varepsilon_1}{SNR \varepsilon_1+SNR^\rho \varepsilon_2 \gamma_s} ) }{\log(SNR)} \Big] \\
 & = \left\{
 \begin{array}{ll}
 \displaystyle
 0 \hspace{3mm} \mbox{If}\ \rho \leq 1
 \vspace{3mm}\\
 \displaystyle
 \rho-1 \hspace{3mm} \mbox{If}\ 1 < \rho \leq 2
 \vspace{3mm}\\
 \displaystyle
 1 \hspace{3mm} \mbox{If}\ \rho > 2
 \end{array}
 \right..
\end{align}
%
%It is easy to verify that the maximum outage diversity of the system is less than 1 since there is only one transmission and reception antenna. Therefore, our scheme can achieve the maximum diversity when $\rho \geq 2$.

By comparing the upper bound on G-SDG in \eqref{eq:cs_diversity_upperbound} and the achievable G-SDG in \eqref{eq:achievable_ol_diversity}, we
have the following theorem.

\begin{theorem}
The modulo-and-forward scheme {achieves} the full generalized
secrecy diversity gain for the untrusted relay channel.
\end{theorem}

%\left\{
% \begin{array}{ll}
% \displaystyle
% 0 ,\hspace{3mm} \mbox{If}\ t\leq1
% \vspace{3mm}\\
% \displaystyle
% \frac{1}{2}\log(1+t/2)-\frac{1}{2}\log(1+t/(t+1))
% \hspace{3mm} \mbox{If}\ 1 < t \leq 3/2
% \vspace{3mm}\\
% \displaystyle
% \frac{1}{2}\log\Big(1+\frac{8t+7}{(2t+1)^2}\Big)
% \hspace{3mm} \mbox{If}\ t > 3/2
% \end{array}
% \right.

\subsection{Comparison with AF Scheme}
This section compares the outage performance of the AF scheme with
that of the MF scheme\footnote{The DF scheme is not compared since
it cannot be adopted without CSIT.}. Since the secrecy outage
probability is independent of the forwarding strategy at the relay,
we need only to calculate the connection outage probability of the
AF scheme, $\poAF$. From the received end-to-end SNR expression for
the AF relaying, we can derive the connection outage probability as
\begin{align}
 \label{eq:AppII:eq1}
 \poAF&=\Pr\Big(\frac{P_s^2 g_1 g_2}{\sigma^2[P_s g_1+(P_s+P_d)g_2+\sigma^2]}<\gamma_{o}\Big)
 \\
 \label{eq:AppII:eq2}
 &=1-e^{-\frac{\gamma_{o}\sigma^2}{P_s}\big(\frac{(P_s+P_d)}{P_s \varepsilon_1}+\frac{1}{\varepsilon_2}\big)}
 \frac{2\gamma_{o}\sigma^2}{P_s}\sqrt{\frac{1}{\varepsilon_1\varepsilon_2}\Big(\frac{P_s+P_d}{P_s}+\frac{1}{\gamma_{o}}\Big)}
 K_1
 \Bigg(\frac{2\gamma_{o}\sigma^2}{P_s}\sqrt{\frac{1}{\varepsilon_1\varepsilon_2}\Big(\frac{P_s+P_d}{P_s}+\frac{1}{\gamma_{o}}\Big)}\Bigg),
\end{align}
where the derivation from (\ref{eq:AppII:eq1}) to
(\ref{eq:AppII:eq2}) is provided in Appendix II.

In Fig. \ref{ol_probability_label}, we plot the outage probability
performance of the AF scheme and the MF scheme, where we set
$P_s=P_d=10$, $\varepsilon_1=\varepsilon_2=\sigma^2=1$, $R_s=1/2$,
and $R_d$ varies from 1/2 to 15. We can see that there is a tradeoff
between the connection outage and secrecy outage probabilities for
both AF and MF schemes. However, the connection outage probability
of the MF scheme is almost half that of the AF scheme with the same
$R_d$, indicating that MF scheme is much more attractive for data
transmission.

\begin{figure}
\begin{center}
\includegraphics[width=5.0in]{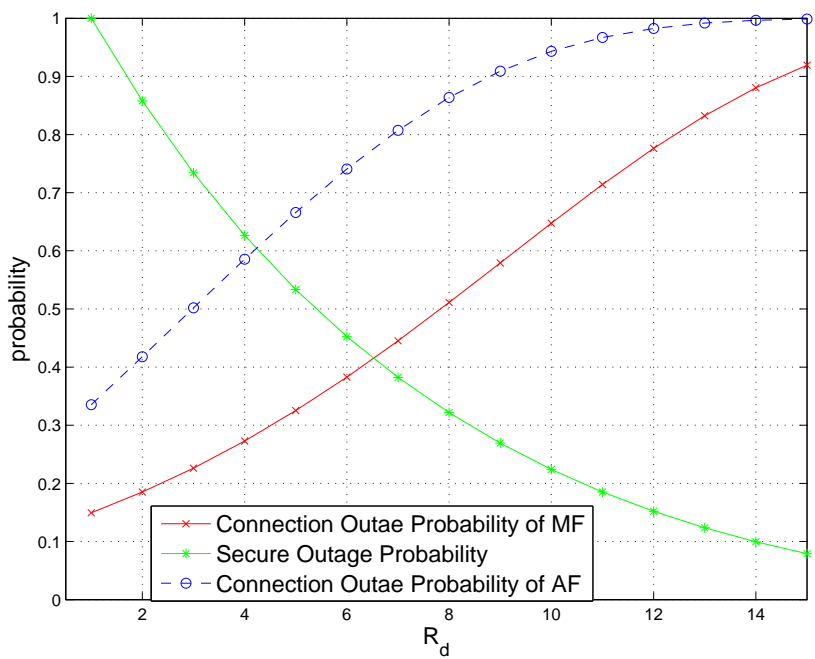}
\caption{Outage probabilities with MF and AF forwarding.} \label{ol_probability_label}
\end{center}
\end{figure}

\begin{figure}
\begin{center}
\includegraphics[width=5.0in]{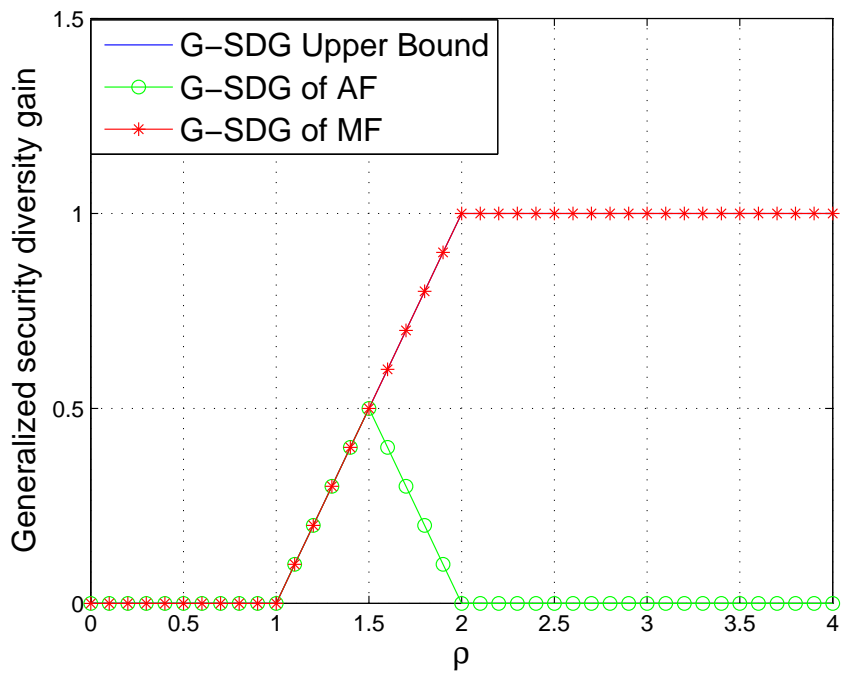}
\caption{Generalized secure diversity gain of AF and MF scheme versus $\rho$.} \label{diversity_gain_label}
\end{center}
\end{figure}

Secondly, we calculate the outage diversity gain of the AF scheme. When the
SNR goes to infinity, $\poAF$ goes to zero only when $\rho \leq
2$. In this case, the asymptotic expression can be written as
$$\poAF\simeq\frac{\gamma_{o}\sigma^2}{P_s}\left(\frac{(P_s+P_d)}{P_s
\varepsilon_1}+\frac{1}{\varepsilon_2}\right).$$ Similar to
\eqref{eq:achievable_ol_diversity}, we can calculate the outage diversity of the AF
scheme as
\begin{align} \label{eq:ol_diversity_af}
 DG_{af}(\rho) & = - \mathop{\lim \sup}_{SNR \to \infty } \frac{-\log ( \max [ \poAF(SNR,\rho), \pl(SNR,\rho)] )}{\log(SNR)} \\
 & = \min \Big[\mathop{\lim \sup}_{SNR \to \infty } \frac{-\log \left(\frac{\gamma_{o}}{SNR}\big(\frac{(SNR+SNR^\rho)}{SNR \varepsilon_1}+\frac{1}{\varepsilon_2}\big)\right) }{\log SNR}, \mathop{\lim \sup}_{SNR \to \infty } \frac{-\log ( \frac{SNR \varepsilon_1}{SNR \varepsilon_1+SNR^\rho \varepsilon_2 \gamma_s} ) }{\log(SNR)} \Big] \\
 & = \left\{
 \begin{array}{ll}
 \displaystyle
 0 \hspace{3mm} \mbox{If}\ \rho \leq 1
 \vspace{3mm}\\
 \displaystyle
 \rho-1 \hspace{3mm} \mbox{If}\ 1 < \rho \leq 1.5
 \vspace{3mm}\\
 \displaystyle
 2-\rho \hspace{3mm} \mbox{If}\ 1.5 < \rho \leq 2
 \vspace{3mm}\\
 \displaystyle
 0 \hspace{3mm} \mbox{If}\ \rho \geq 2
 \end{array}
 \right..
\end{align}
Obviously, the maximum outage diversity of the AF scheme is $1/2$,
which is only half of that of the MF scheme. Moreover, this
diversity of $1/2$ is achieved only when $\rho=3/2$, which requires
very strict power control at the destination. These phenomena can be
numerically observed in Fig. \ref{diversity_gain_label}, where a
comparison of generalized secure diversity gain between the the AF
and the MF schemes is plotted.

\section{Conclusion}
In this paper, we have considered the untrusted relay channel. Inspired by the MLAN channel, we have proposed a modulo
operation before forwarding at the relay, at which a lattice
encoded confidential message from the source and  Gaussian
distributed artificial noise from the destination are superimposed. As a result, the artificial noise only
helps to protect the confidential message from eavesdropping at the
untrusted relay, with almost no detrimental effect of wasted relay
power. For the case with full CSIT, we have shown that the proposed
MF scheme approaches the secrecy capacity within $1/2$ bit for any
channel realization, hence achieving full generalized secure degrees of
freedom. For the case without  CIST, we have shown that the
proposed MF scheme achieves the full generalized secure diversity
gain of $1$. On the other hand, the traditional AF scheme only
achieves a G-SDG of $1/2$ at most.

\appendices
\section{Derivation of Eq. (\ref{eq:App:eq2})}
\label{App:A}
\renewcommand{\theequation}{A.\arabic{equation}}
\setcounter{equation}{0}

From (\ref{eq:App:eq1}), we can write the achievable connection
outage probability as
\begin{align}
 \po&=\Pr\Big[\frac{P_s}{\sigma^2}\frac{g_1
 g_2}{g_1+g_2}<\gamma_{1}\Big],
\end{align}
where $\gamma_{1}=2^{2R_d}-1/2$ is the target SNR threshold. We can
further write $p_{o}$ as
\begin{align}
 \po&=\Pr\Big(g_1 g_2<\frac{\gamma_{1}\sigma^2}{P_s} g_1+\frac{\gamma_{1}\sigma^2}{P_s}g_2\Big) \notag
 \\
 &=\Pr\Big[g_1(g_2-\frac{\gamma_{1}\sigma^2}{P_s})<\frac{\gamma_{1}\sigma^2}{P_s} g_2\Big].
\end{align}
Considering the two cases of $g_2\leq
\frac{\gamma_{1}\sigma^2}{P_s}$ and
$g_2>\frac{\gamma_{1}\sigma^2}{P_s}$, we can rewrite $\po$ as
\begin{align}
 \po&=\Pr\Big(g_2\leq\frac{\gamma_{1}\sigma^2}{P_s}\Big)
 +\Pr\Big(g_2>\frac{\gamma_{1}\sigma^2}{P_s},g_1<\frac{\frac{\gamma_{1}\sigma^2}{P_s} g_2}{g_2-\frac{\gamma_{1}\sigma^2}{P_s}}\Big).
\end{align}
Applying the probability density functions
$f_{g_1}(x)=\frac{1}{\varepsilon_1}e^{-\frac{x}{\varepsilon_1}}$ and
$f_{g_2}(x)=\frac{1}{\varepsilon_2}e^{-\frac{x}{\varepsilon_2}}$ in
the above equation, we can obtain an analytical expression for the
outage probability
 as
\begin{align} %\label{eq:outage_probability}
 \po &=\Pr\Big(g_2\leq\frac{\gamma_{1}\sigma^2}{P_s}\Big)+\int_{0}^{\frac{\frac{\gamma_{1}\sigma^2}{P_s} g_2}{g_2-\frac{\gamma_{1}\sigma^2}{P_s}}}
 f_{g_1}(g_1)\Pr\Big(g_2>\frac{\gamma_{1}\sigma^2}{P_s}\Big)dg_1 \notag
 \\&
 =1-\frac{1}{\varepsilon_2}\int_{g_2=\frac{\gamma_{1}\sigma^2}{P_s}}e^{-\Big(\frac{1}{\varepsilon_2}+
 \frac{\gamma_{1}\sigma^2}{P_s\varepsilon_1(g_2-\frac{\gamma_{1}\sigma^2}{P_s})}\Big)g_2}dg_2
 \\&
 =1-e^{-(\frac{1}{\varepsilon_1}+\frac{1}{\varepsilon_2})\frac{\gamma_{1}\sigma^2}{P_s}}
 \frac{2\gamma_{1}\sigma^2}{P_s \sqrt{\varepsilon_1
 \varepsilon_2}}K_1\Big( \frac{2\gamma_{1}\sigma^2}{P_s \sqrt{\varepsilon_1
 \varepsilon_2}}\Big).
\end{align}
Thus, we have (\ref{eq:App:eq2}).

\section{Derivation of Eq. (\ref{eq:AppII:eq2})}
\label{App:B}
\renewcommand{\theequation}{B.\arabic{equation}}
\setcounter{equation}{0}

From (\ref{eq:AppII:eq1}), we can write the connection outage
probability of AF relaying as
\begin{align}
 \poAF&=\Pr\big(P_s^2 g_1 g_2<P_s \sigma^2 \gamma_{o}g_1+(P_s+P_d)\sigma^2\gamma_{o}g_2+\gamma_{o}\sigma^4 \big)
 \\&=\Pr\big[(P_s^2 g_2-P_s
 \sigma^2\gamma_{o})g_1<(P_s+P_d)\sigma^2\gamma_{o}g_2+\gamma_{o}\sigma^4\big].
\end{align}
Considering the two cases of $g_2\leq
\frac{\gamma_{o}\sigma^2}{P_s}$ and $g_2>
\frac{\gamma_{o}\sigma^2}{P_s}$ separately, we can further write
$\poAF$ as
\begin{align}
 \poAF&=\Pr(g_2\leq
 \frac{\gamma_{o}\sigma^2}{P_s})+\Pr\Big(g_2> \frac{\gamma_{o}\sigma^2}{P_s},g_1<\frac{\gamma_{o}\sigma^2}{P_s}
 \cdot \frac{(P_s+P_d)g_2+\sigma^2}{P_s g_2-\gamma_{o}\sigma^2}\Big)
 \\
 &=1-\frac{1}{\varepsilon_2}\int_{\frac{\gamma_{o}\sigma^2}{P_s}}^\infty
 e^{-\frac{g_2}{\varepsilon_2}-\frac{\gamma_{o}\sigma^2}{P_s \varepsilon_1}
 \cdot \frac{(P_s+P_d)g_2+\sigma^2}{P_s
 g_2-\gamma_{o}\sigma^2}}dg_2.
\end{align}
Using the variable substitution of
$x=g_2-\frac{\gamma_{o}\sigma^2}{P_s}$, we can obtain a
closed-form expression for the connection outage probability for the
AF relaying as
\begin{align}
 \poAF&=1-e^{-\frac{\gamma_{o}\sigma^2}{P_s}\big(\frac{(P_s+P_d)}{P_s \varepsilon_1}+\frac{1}{\varepsilon_2}\big)}
 \frac{2\gamma_{o}\sigma^2}{P_s}\sqrt{\frac{1}{\varepsilon_1\varepsilon_2}\Big(\frac{P_s+P_d}{P_s}+\frac{1}{\gamma_{o}}\Big)}
 K_1
 \Bigg(\frac{2\gamma_{o}\sigma^2}{P_s}\sqrt{\frac{1}{\varepsilon_1\varepsilon_2}\Big(\frac{P_s+P_d}{P_s}+\frac{1}{\gamma_{o}}\Big)}\Bigg).
\end{align}
In this way, the proof of (\ref{eq:AppII:eq2}) has been
completed.

\bibliographystyle{IEEEtran}
\bibliography{./final_refs}

\end{document}